\documentclass[10pt,twocolumn,twoside]{IEEEtran}
\usepackage{calc}

\usepackage{multirow}
\usepackage{cite}
\usepackage{graphicx,subfigure}
\usepackage{psfrag}
\usepackage{amsmath,amssymb}
\usepackage{color}

\usepackage{amsmath,amssymb,eucal}
\usepackage{xifthen}
\usepackage{mathtools}
\usepackage{enumerate}
\usepackage{microtype}
\usepackage{xspace}
\usepackage{bm}
\usepackage[T1]{fontenc}
\usepackage{fancyhdr}
\usepackage{lastpage}

\newcommand{\mbs}[1]{\bm{#1}}
\newcommand{\vect}[1]{{\lowercase{\mbs{#1}}}}

\newcommand{\T}{{\scriptscriptstyle\mathsf{T}}}
\renewcommand{\H}{{\scriptscriptstyle\mathsf{H}}}

\renewcommand{\Re}[1][]{\ifthenelse{\isempty{#1}}{\operatorname{Re}}{\operatorname{Re}\left(#1\right)}}
\renewcommand{\Im}[1][]{\ifthenelse{\isempty{#1}}{\operatorname{Im}}{\operatorname{Im}\left(#1\right)}}

\newcommand{\gv}{\vect{g}}
\newcommand{\hv}{\vect{h}}

\newcommand{\uv}{\vect{u}}
\newcommand{\vv}{\vect{v}}
\newcommand{\wv}{\vect{w}}
\newcommand{\xv}{\vect{x}}

\newcommand{\Xc}{{\mathcal X}}

\newcommand{\CC}{\mathbb{C}}

\newcommand{\CN}[1][]{\ifthenelse{\isempty{#1}}{\mathcal{N}_{\mathbb{C}}}{\mathcal{N}_{\mathbb{C}}\left(#1\right)}}

\renewcommand{\P}[1][]{\ifthenelse{\isempty{#1}}{\mathbb{P}}{\mathbb{P}\left(#1\right)}}
\newcommand{\E}[1][]{\ifthenelse{\isempty{#1}}{\mathbb{E}}{\mathbb{E}\left(#1\right)}}
\renewcommand{\det}[1][]{\ifthenelse{\isempty{#1}}{\text{det}}{\text{det}\left(#1\right)}}
\newcommand{\trace}[1][]{\ifthenelse{\isempty{#1}}{\text{tr}}{\text{tr}\left(#1\right)}}
\newcommand{\rank}[1][]{\ifthenelse{\isempty{#1}}{\text{rank}}{\text{rank}\left(#1\right)}}
\newcommand{\diag}[1][]{\ifthenelse{\isempty{#1}}{\text{diag}}{\text{diag}\left(#1\right)}}

\DeclarePairedDelimiter\norm{\lVert}{\rVert}

\newcommand{\defeq}{\triangleq}

\newtheorem{remark}{Remark}[section]
\newtheorem{theorem}{Theorem}%[section]
\newtheorem{corollary}{Corollary}%[section]
\newcounter{enumi_saved}
\usepackage{answers}
\Newassociation{solution}{Solution}{solutionfile}

\AtBeginDocument{\Opensolutionfile{solutionfile}[\jobname]}
\AtEndDocument{\Closesolutionfile{solutionfile}\clearpage
}

\newcommand{\bit}{\begin{itemize}}
\newcommand{\eit}{\end{itemize}}

\newcommand{\ba}{\begin{array}}
\newcommand{\ea}{\end{array}}

\newcommand{\beq}{\begin{equation}}
\newcommand{\eeq}{\end{equation}}

\newcommand{\beqn}{\begin{equation*}}
\newcommand{\eeqn}{\end{equation*}}

\newcommand{\bean}{\begin{eqnarray*}}
\newcommand{\eean}{\end{eqnarray*}}
\newcommand{\bea}{\begin{eqnarray}}
\newcommand{\eea}{\end{eqnarray}}

\IfFileExists{MinionPro.sty}{
}{
}
\def\bit{\begin{itemize}}
\def\eit{\end{itemize}}

\begin{document}
\sloppy

\title{Imperfect Delayed CSIT can be as Useful as Perfect Delayed CSIT: DoF Analysis and Constructions for the BC}

\author{Jinyuan Chen and Petros Elia
\thanks{The research leading to these results has received funding from the European Research Council under the European Community's Seventh Framework Programme (FP7/2007-2013) / ERC grant agreement no. 257616 (CONECT), from the FP7 CELTIC SPECTRA project, and from Agence Nationale de la Recherche project ANR-IMAGENET.
}
\thanks{J. Chen and P. Elia are with the Mobile Communications Department, EURECOM, Sophia Antipolis, France (email: \{chenji, elia\}@eurecom.fr)}
}

%%%%%%%%%%%%%%%%%%%%%%%%%%%%%%%%%%%%%%%%%%%%%

\maketitle
\thispagestyle{empty}

%%%%%%%%%%%%%%%%%%%%%%%%%%%%%%%%%%%%%%%%%%%%%
\begin{abstract}
In the setting of the two-user broadcast channel, where a two-antenna transmitter communicates information to two single-antenna receivers, recent work by Maddah-Ali and Tse has shown that perfect knowledge of delayed channel state information at the transmitter (perfect delayed CSIT) can be useful, even in the absence of any knowledge of current CSIT.  Similar benefits of perfect delayed CSIT were revealed in recent work by Kobayashi et al., Yang et al., and Gou and Jafar, which extended the above to the case of perfect delayed CSIT and imperfect current CSIT.

The work here considers the general problem of communicating, over the aforementioned broadcast channel, with imperfect delayed and imperfect current CSIT, and reveals that even substantially degraded and imperfect delayed-CSIT is in fact sufficient to achieve the aforementioned gains previously associated to perfect delayed CSIT.  The work proposes novel multi-phase broadcasting schemes that properly utilize knowledge of imperfect delayed and imperfect current CSIT, to match in many cases the optimal degrees-of-freedom (DoF) region achieved with perfect delayed CSIT.
In addition to the theoretical limits and explicitly constructed precoders, the work applies towards gaining practical insight as to when it is worth improving CSIT quality.
\end{abstract}

%%%%%%%%%%%%%%%%%%%%%%%%%%%%%%%%%%%%%%%%%%%
\section{Introduction}
In many multiuser wireless communications scenarios, having sufficient CSIT is a crucial ingredient that facilitates improved performance.  While being useful, perfect CSIT is also hard and time-consuming to obtain, hence the need for communication schemes that can utilize imperfect and delayed CSIT knowledge {(\cite{MAT:11c,GMK:11o,VV:10t,AGK:11o,GMK:11i,XAJ:11b})}.  In this context of multiuser communications, we here consider the broadcast channel (BC), and specifically focus on the two-user multiple-input single-output (MISO) BC, where a two-antenna transmitter communicates information to two single-antenna receivers.  In this setting, the channel model takes the form

\begin{subequations}
\begin{align}
y^{(1)}_t &= \hv^{\T}_t \xv_t + z^{(1)}_t      \label{eq:modely1}\\
y^{(2)}_t &= \gv^{\T}_t \xv_t + z^{(2)}_t,      \label{eq:modely2}
\end{align}
\end{subequations}
where for any time instance $t$, vectors $\hv_t, \gv_t \in \CC^{2\times 1}$ represent
the transmitter-to-user~1 and transmitter-to-user~2 channels respectively, where $z^{(1)}_t,z^{(2)}_t$ represent unit power AWGN noise at the two receivers, where $\xv_t$ is the input signal with power
constraint $\E[ \norm{\xv_t}^2 ] \le P$, and where in this case, $P$ also takes the role of the signal-to-noise ratio (SNR).

With CSIT often being imperfect and delayed, we here explore the effects of the \emph{quality of current CSIT} corresponding to how well the transmitter knows $\hv_{t},\gv_{t}$ at time $t$, as well as the effects of the \emph{quality of delayed CSIT}, corresponding to how well the transmitter knows the same $\hv_{t},\gv_{t}$, at time $t+\tau$ for some positive $\tau$.
Naturally, reduced CSIT quality relates to limitations in the capacity and reliability of the feedback channel. The distinction between the quality of current and delayed CSIT, is meant to reflect the increased challenge of quickly attaining high quality CSIT.

\subsection{Related work}
Corresponding to CSIT quality, it is well known that in the two-user BC setting of interest, the presence of perfect CSIT allows for the optimal $1$ degree-of-freedom (DoF) per user, whereas the complete absence of CSIT causes a substantial degradation to just $1/2$ DoF per user\footnote{We remind the reader that for an achievable rate pair $(R_1,R_2)$, the corresponding DoF pair $(d_1,d_2)$ is given by $d_i = \lim_{P \to \infty} \frac{R_i}{\log P},\ i=1,2.$  The corresponding DoF region is then the set of all achievable DoF pairs.}.

An interesting scheme utilizing partial CSIT knowledge, was recently presented in \cite{MAT:11c} by Maddah-Ali and Tse, which showed that delayed CSIT knowledge can still be useful in improving the DoF region of the broadcast channel.  In the above described two-user MISO BC setting, and under the assumption that at time $t$, the transmitter perfectly knows the delayed channel states ($\hv,\gv$) up to time $t-1$ (perfect delayed, no current CSIT), the work in~\cite{MAT:11c} showed that each user can achieve $2/3$ DoF, providing a clear improvement over the case of no CSIT.
This result was later generalized in \cite{MJS:12r,KYGY:12o,YKGY:12d,GJ:12o,CE:12d} which considered the natural extension where, in addition to perfect delayed CSIT, the transmitter also had partial knowledge of current CSIT.

\subsection{Notation and conventions}
Throughout this paper, $(\bullet)^\T$, $(\bullet)^{\H}$, respectively denote the transpose and conjugate transpose of a matrix, while $||\bullet||$ denotes the Euclidean norm, and $|\bullet|$ denotes the magnitude of a scalar.
$o(\bullet)$ comes from the standard Landau notation, where $f(x) = o(g(x))$ implies $\lim_{x\to \infty} f(x)/g(x)=0$.  We also use $\doteq$ to denote \emph{exponential equality}, i.e., we write $f(P)\doteq P^{B}$ to denote $\displaystyle\lim_{P\to\infty}\frac{\log f(P)}{\log P}=B$.  Logarithms are of base~$2$.

Finally adhering to the convention followed in \cite{MAT:11c,MJS:12r,GJ:12o}, we consider a unit coherence period\footnote{Simple interleaving arguments can show that, in the absence of delay constraints, the association of current CSIT with a single coherence period, introduces no loss of generality.}, as well as perfect and global knowledge of channel state information at the receivers (perfect global CSIR, \cite{MAT:11c,MJS:12r,GJ:12o,YKGY:12d}) where the receivers know all channel states and all estimates.

\subsection{Structure of paper}
After recalling the quantification of CSIT quality, Section~\ref{sec:model-results} bounds the DoF region of the described two-user MISO broadcast channel for the general case of having imperfect current and imperfect delayed CSIT of different quality. In many cases, these bounds are identified to be tight, and to in fact match the optimal performance associated to perfect delayed CSIT.  Section~\ref{sec:achievability} presents the novel multi-phase precoding schemes that apply for different cases of CSIT quality.  The performance of these schemes is derived in the same section, with some of the proof details placed in the Appendix.

%=====================================
\section{MISO BC with imperfect delayed CSIT and imperfect current CSIT\label{sec:model-results}}
\subsection{Quantification of CSIT quality}
In terms of current CSIT, we consider the case where at time $t$, the transmitter has estimates $\hat{\hv}_t,\hat{\gv}_t$ of $\hv_t$ and $\gv_t$ respectively, with estimation errors
\begin{equation}
\label{eq:MMSEc}
  \tilde{\hv}_t  = \hv_t - \hat{\hv}_t, \ \ \     \tilde{\gv}_t  = \gv_t - \hat{\gv}_t
\end{equation}
having i.i.d. Gaussian entries with power \[\frac{1}{2}\E[\|\tilde{\hv}_t\|^2] =\frac{1}{2}\E[\|\tilde{\gv}_t\|^2] = P^{-\alpha},\] for some non-negative parameter $\alpha$ describing the quality of the estimates.  In this setting, an increasing $\alpha$ implies an improved CSIT quality, with $\alpha = 0$ implying very little current CSIT knowledge, and with $\alpha = \infty$ implying perfect CSIT.

In terms of delayed CSIT for channels $\hv_{t},\gv_{t}$ that appear at time $t$, we consider the case where, beginning at time $t+1$, the transmitter has delayed estimates $\check{\hv}_{t},\check{\gv}_{t}$ of $\hv_{t},\gv_{t}$, and does so with estimation errors
\begin{equation}
\label{eq:MMSEd}
  \ddot{\hv}_{t}  = \hv_{t} - \check{\hv}_{t}, \ \ \     \ddot{\gv}_{t}  = \gv_{t} - \check{\gv}_{t}
\end{equation}
having i.i.d. Gaussian entries with power \[\frac{1}{2}\E[\|\ddot{\hv}_{t}\|^2] =\frac{1}{2}\E[\|\ddot{\gv}_{t}\|^2]=P^{-\beta},\] for some non-negative parameter $\beta$ describing the quality of the estimates.

\begin{remark}
We here note that without loss of generality, we can restrict our attention to the range $0\leq \alpha,\beta\leq 1$ (cf.\cite{Caire+:10m}), as well as to the case where $\alpha\leq \beta$ since having $\alpha >\beta$ would be equivalent to having $\alpha = \beta$ simply because current CSIT estimates can be recalled at a later time.  As a result, we will henceforth consider that $0\leq \alpha \leq  \beta \leq 1,$ where $\beta = 1$ corresponds the case of perfect delayed CSIT, and where $\alpha = 1$ corresponds to the case of perfect CSIT.
\end{remark}

\begin{figure}
	\centering
	\includegraphics[width = 8.8cm]{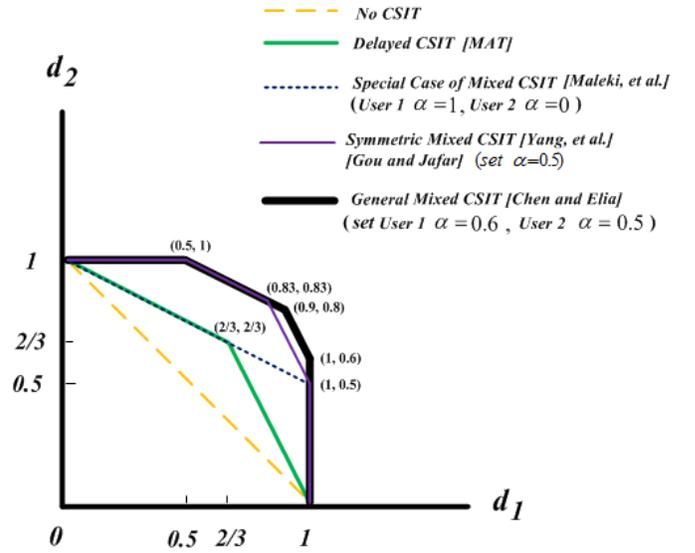}
	\caption{DoF regions: Imperfect current CSIT ($0\leq \alpha\leq 1$), and perfect delayed CSIT ($\beta = 1$).}
	\label{fig:DoFR_HistoryMixedCSITPerfectDCSIT}
\end{figure}

Fig.~\ref{fig:DoFR_HistoryMixedCSITPerfectDCSIT} recalls different DoF regions corresponding to imperfect current CSIT ($0\leq \alpha\leq 1$), but perfect delayed CSIT ($\beta = 1$) (\cite{YKGY:12d,GJ:12o,MJS:12r,CE:12d}).

%=======================================
\subsection{DoF region of the MISO BC with imperfect delayed and imperfect current CSIT \label{sec:bc-dof}}

We proceed with the main result, the proof of which, together with the description of the associated precoding schemes, will be given in Section~\ref{sec:achievability}.

\vspace{3pt}
\begin{theorem} \label{theorem:bc-iie}
For the two-user MISO BC with imperfect delayed CSIT, imperfect current CSIT $(0\leq\alpha\leq \beta\leq 1)$, and for $\beta^{''}\defeq \min\{\beta, \frac{1+2\alpha}{3}\}$, the DoF region
  \begin{align}
	   d_1 \le 1, \quad \ \  \  d_2 \le 1  \nonumber\\%\label{eq:them-iie1}\\
     (1+\beta^{''}-2\alpha)d_1  + (1-\beta^{''}) d_2 \le (1+\beta^{''})(1-\alpha) \nonumber\\%\label{eq:them-iie2}\\
     (1-\beta^{''}) d_1  + (1+\beta^{''}-2\alpha)d_2 \le (1+\beta^{''})(1-\alpha) \nonumber%\label{eq:them-iie3}
  \end{align}
is achievable and takes the form of a polygon with corner points \[\{(0,0),(0,1),(\alpha,1),(\frac{1+\beta^{''}}{2},\frac{1+\beta^{''}}{2}),(1,\alpha),(1, 0)\}.\]
Furthermore when $\beta\geq\frac{1+2\alpha}{3}$, the region is optimal and it is described by
  \begin{align*}
	   d_1 \le 1, \quad  d_2 \le 1  \nonumber\\
     2d_1  + d_2 \le 2+\alpha \nonumber\\
     d_1  + 2 d_2 \le 2+\alpha  \nonumber
  \end{align*}
corresponding to the polygon \[\{(0,0), (0,1), (\alpha,1), (\frac{2+\alpha}{3},\frac{2+\alpha}{3}),(1,\alpha),(1, 0)\}\] matching the optimal DoF region previously associated to $\beta~=~1$.
\end{theorem}
\vspace{3pt}

The above reveals that, whether with imperfect or no current CSIT, imperfect delayed CSIT can in some cases match the optimal performance associated to perfect delayed CSIT.  The following corollaries provide further insight, and make the connection to previous work.  The corollaries apply to the same setting as the theorem.
\begin{corollary}
In terms of DoF, having $\beta\geq \frac{1+2\alpha}{3}$ is equivalent to having perfect delayed CSIT.
Specifically the optimal region $\{(0,0), (0,1), (\alpha,1), (\frac{2+\alpha}{3},\frac{2+\alpha}{3}),(1,\alpha),(1, 0)\}$ from \cite{YKGY:12d,GJ:12o} corresponding to $\beta~=~1$, can in fact be achieved for any $\beta\geq \frac{1+2\alpha}{3}$, and the optimal region $\{(0,0), (0,1), (\frac{2}{3},\frac{2}{3}),(1, 0)\}$ from~\cite{MAT:11c} corresponding to $\beta = 1,\alpha = 0$, can in fact be achieved whenever $\beta\geq 1/3$.
\end{corollary}
\vspace{3pt}

Building on the above, we also have the following.
\begin{corollary}
Whenever the desired DoF pair lies within the pentagon $\{(0,0), (0,1), (\alpha,1),(1,\alpha),(1, 0)\}$, there is no need for any delayed CSIT, and $\beta = 0$ suffices.
\end{corollary}
\vspace{3pt}

This is the case for example, for the optimal $d_1 = 1, d_2 = \alpha$, which can be achieved with imperfect current and no delayed CSIT.  Consequently whenever the desired DoF pair lies within the aforementioned pentagon, or whenever $\beta\geq \frac{1+2\alpha}{3}$, then there is no need for improving the quality of delayed CSIT.  Otherwise, the DoF penalty due to a reduced $\beta$, can be at most $\frac{2+\alpha}{3} - \frac{1+\beta^{''}}{2} = \frac{1+2\alpha-3\beta}{6}$, which is no bigger than $\frac{1-\alpha}{6}$.

\begin{figure}
	\centering
	\includegraphics[width = 8cm]{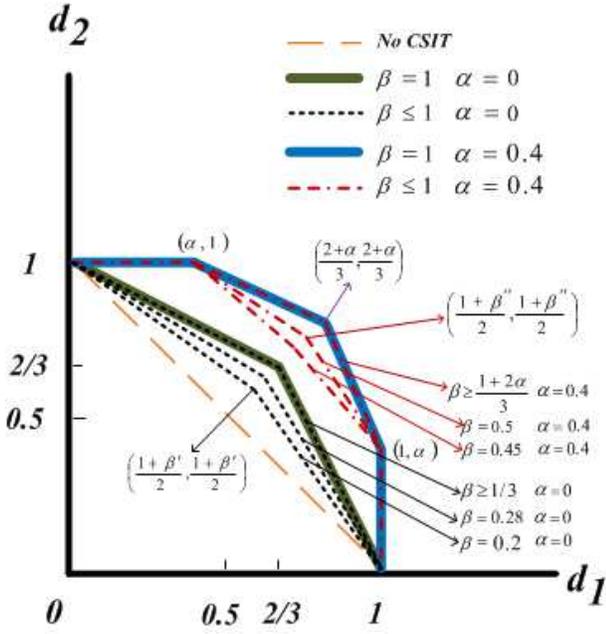}
	\caption{DoF regions with imperfect current CSIT and imperfect delayed CSIT.  Recall $\beta^{'}=\min\{\beta, \frac{1}{3}\}$ and $\beta^{''}=\min\{\beta, \frac{1+2\alpha}{3}\}$.}
	\label{fig:DoFR_HistoryImperfectDCSIT}
\end{figure}

Fig.~\ref{fig:DoFR_HistoryImperfectDCSIT} depicts different DoF regions spanning the general setting of imperfect delayed and imperfect current CSIT.

\begin{figure}
\centering
\includegraphics[width=8.5cm]{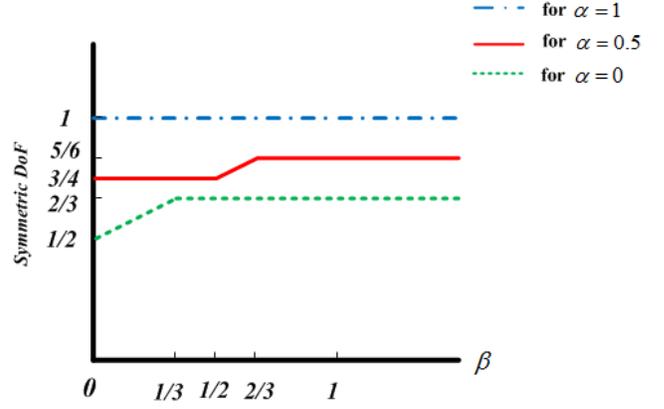}
\caption{Achievable symmetric DoF ($0\leq \beta \leq 1, \alpha=0,0.5,1$).}
\label{fig:DoFX1X2}
\end{figure}

%=====================================
\section{Multi-phase precoding schemes for the two-user MISO BC with imperfect delayed and imperfect current CSIT\label{sec:achievability}}

We proceed to describe the two precoding schemes that achieve the corresponding corner DoF points, % (see Table~\ref{tab:allsummary}),
by properly utilizing different combinations of superposition coding, successive cancelation, power allocation, and phase durations.

As stated, without loss of generality, we assume that $0\leq \alpha \leq  \beta \leq 1$.
The scheme description is done for $0 < \alpha < \beta < 1$, and for rational $\alpha, \beta$.  The cases where $\beta=1$, $\beta=\alpha$, $\alpha=0$, or where $\alpha,\beta$ are not rational, can be readily handled with minor modifications.
We first proceed to describe the basic notation and conventions used in our schemes.  This preliminary description allows for brevity in the subsequent description of the details of the schemes.

The schemes are designed with $S$ phases ($S$ varies from scheme to scheme), where the $s$th phase ($s=1,2,\cdots,S$) consists of $T_s$ channel uses.  At this point, and to more clearly reflect the division of time into phases, we will switch to a double time index where, for example, the vectors $\hv_{s,t}$ and $\gv_{s,t}$ will now denote the channel vectors, during timeslot $t$ of phase $s$.  Similarly, in terms of current CSIT (cf.~\eqref{eq:MMSEc}), $\hat{\hv}_{s,t}$ and $\hat{\gv}_{s,t}$ will respectively denote the transmitter's estimates of channels $\hv_{s,t}$ and $\gv_{s,t}$, and $\tilde{\hv}_{s,t} = \hv_{s,t} - \hat{\hv}_{s,t} $, $\tilde{\gv}_{s,t} = \gv_{s,t} - \hat{\gv}_{s,t} $ will denote the corresponding estimation errors.  We recall that the estimates $\hat{\hv}_{s,t}$ and $\hat{\gv}_{s,t}$ become known to the transmitter at time $t$, i.e., they become known instantly.
In terms of delayed CSIT (cf.~\eqref{eq:MMSEd}), $\check{\hv}_{s,t}$ and $\check{\gv}_{s,t}$ will be the estimates of $\hv_{s,t}$ and $\gv_{s,t}$, where these estimates become known to the transmitter with unit delay (at time $t+1$), and are stored and recalled thereafter.  Finally $\ddot{\hv}_{s,t} = \hv_{s,t} - \check{\hv}_{s,t} $, $\ddot{\gv}_{s,t} = \gv_{s,t} - \check{\gv}_{s,t} $ will denote the estimation errors corresponding to delayed CSIT.

Furthermore $a_{s,t}$ and $a^{'}_{s,t}$ will denote the independent information symbols that are precoded and sent during phase~$s$, timeslot~$t$, and which are meant for user 1, while symbols $b_{s,t}$ and $b^{'}_{s,t}$ are meant for user 2. In addition, $c_{s,t}$ will denote the common information symbol generally meant for both users.

The transmitted vector at timeslot $t$ of phase $s$ will, in most cases, take the form
\beq\label{eq:TxGeneralg}
\xv_{s,t} =\wv_{s,t} \underbrace{c_{s,t}}_{P_s^{(c)}} +\uv_{s,t} \underbrace{a_{s,t}}_{P_s^{(a)}} + \uv^{'}_{s,t} \underbrace{a^{'}_{s,t}}_{P_s^{(a')}}+\vv_{s,t}\underbrace{b_{s,t}}_{P_s^{(b)}} +\vv^{'}_{s,t} \underbrace{b^{'}_{s,t}}_{P_s^{(b')}},
\eeq
where vectors $\wv_{s,t}, \uv_{s,t},\uv^{'}_{s,t}, \vv_{s,t},\vv^{'}_{s,t}$ are the unit-norm beamformers for $c_{s,t},a_{s,t},a^{'}_{s,t}, b_{s,t},b^{'}_{s,t}$ respectively.
In our schemes, vectors $\uv_{s,t}$ and $\vv_{s,t}$ will be chosen to be orthogonal to $\hat{\gv}_{s,t}$ and $\hat{\hv}_{s,t}$ respectively, with $\wv_{s,t},\uv^{'}_{s,t}, \vv^{'}_{s,t}$ chosen pseudo-randomly (and assumed to be known by all nodes).
Corresponding to the transmitted vector in \eqref{eq:TxGeneralg}, and as noted under each summand, the average power that is assigned to each symbol, throughout a specific phase, will be denoted as follows:
\[\begin{array}{ccc} P^{(c)}_s \defeq \E |c_{s,t}|^2, & P^{(a)}_s \defeq \E |a_{s,t}|^2, & P^{(a')}_s \defeq \E |a^{'}_{s,t}|^2 \\ P^{(b)}_s \defeq \E |b_{s,t}|^2, & P^{(b')}_s \defeq \E |b^{'}_{s,t}|^2.\end{array}\]
Furthermore, regarding the amount of information, per time slot, carried by each of the above symbols, we will use $r^{(a)}_s$ to mean that, during phase~$s$, each symbol $a_{s,t},  \ t = 1,\cdots,T_s,$ carries $r^{(a)}_s\log P +o(\log P)$ bits, and similarly we will use $r^{(a')}_s,r^{(b)}_s,r^{(b')}_s,r^{(c)}_s$ to describe the prelog factor of the number of bits in $a^{'}_{s,t}, b_{s,t},b^{'}_{s,t},c_{s,t}$ respectively, again for phase~$s$.

In addition, we will use
\begin{align} \label{eq:c2barsg}
\iota^{(1)}_{s,t} & \defeq \hv^\T_{s,t}(\vv_{s,t} b_{s,t}+\vv^{'}_{s,t} b^{'}_{s,t}), \nonumber\\
\iota^{(2)}_{s,t} & \defeq \gv^\T_{s,t} (\uv_{s,t} a_{s,t} + \uv^{'}_{s,t} a^{'}_{s,t}), \ t = 1,\cdots,T_s
\end{align}
to denote the new interference experienced by user~1 and user~2 respectively, during timeslot $t$ of phase $s$, and we will use
\begin{align} \label{eq:cbarg}
\check{\iota}^{(1)}_{s,t} & \defeq \check{\hv}^\T_{s,t}(\vv_{s,t} b_{s,t}+\vv^{'}_{s,t} b^{'}_{s,t}) , \nonumber\\
\check{\iota}^{(2)}_{s,t} & \defeq \check{\gv}^\T_{s,t} (\uv_{s,t} a_{s,t} + \uv^{'}_{s,t} a^{'}_{s,t}), \quad  t = 1,\cdots,T_s,
\end{align}
to denote transmitter's (delayed) estimates of $\iota^{(1)}_{s,t},\iota^{(2)}_{s,t}$ at time $t+1$.  To clarify, we mean that the transmitter creates, at time $t+1$, the estimates $\check{\iota}^{(2)}_{s,t},\check{\iota}^{(1)}_{s,t}$ of the actual interference $\iota^{(2)}_{s,t},\iota^{(1)}_{s,t}$ experienced during time $s,t$, by using the delayed CSIT estimates obtained at time $t+1$.

For $\{\check{\iota}^{(2)}_{s,t},\check{\iota}^{(1)}_{s,t}\}_{t=1}^{T_{s}}$ being the accumulated delayed estimates of all the interference terms during phase $s$, we will let $\{\bar{\check{\iota}}^{(2)}_{s,t},\bar{\check{\iota}}^{(1)}_{s,t}\}_{t=1}^{T_{s}}$ be the quantized delayed estimates which are obtained by properly quantizing $\{\check{\iota}^{(2)}_{s,t},\check{\iota}^{(1)}_{s,t}\}_{t=1}^{T_{s}}$, at a quantization rate that will be described later on.  Based on the information in $\{\check{\iota}^{(2)}_{s,t},\check{\iota}^{(1)}_{s,t}\}_{t=1}^{T_{s}}$, new symbols $\{c_{s+1,t}\}_{t=1}^{T_{s+1}}$ are then created, where these new symbols are created to evenly share the total information in $\{\bar{\check{\iota}}^{(2)}_{s,t},\bar{\check{\iota}}^{(1)}_{s,t}\}_{t=1}^{T_{s}}$ (i.e., the information in $\{\check{\iota}^{(2)}_{s,t},\check{\iota}^{(1)}_{s,t}\}_{t=1}^{T_{s}}$ is evenly split among the elements in $\{c_{s+1,t}\}_{t=1}^{T_{s+1}}$), and where these new common symbols will be sequentially transmitted during the next phase.

Finally the received signals $y^{(1)}_{s,t}$ and $y^{(2)}_{s,t}$ at the first and second user during phase $s$, take the form
\begin{align}
  y^{(1)}_{s,t}&= \hv^\T_{s,t} x_{s,t}+z^{(1)}_{s,t},\nonumber \\
  y^{(2)}_{s,t}&= \gv^\T_{s,t} x_{s,t}+z^{(2)}_{s,t}, \ t = 1,\cdots, T_s.
\end{align}
%-----------------------------------
%----------X1-----------------------
%-----------------------------------
We now proceed with the details of the first scheme.

\subsection{Scheme $\Xc_1$ achieving $C_1=\frac{1+\beta^{''}}{2}$ ($0 \leq \alpha \leq \beta\leq 1$)}

For this scheme, the phase durations $T_1,T_2,\cdots,T_S$ are chosen to be integers generated to form a geometric progression where
\begin{align}
T_s&=T_{s-1} \xi=T_1\xi^{s-1}, \forall s\in \{2,3,\cdots,S-1\},  \nonumber\\
T_{S}&=T_{S-1}\zeta=T_1\xi^{S-2}\zeta,  \label{eq:X1T}
\end{align}
and where $\xi=\frac{2(\beta-\alpha)}{1-\beta}$, $\zeta=\frac{2(\beta-\alpha)}{1-\alpha}$.  The progression can be made to consist of integers since $\alpha,\beta$, and by extension $\zeta,\xi$, are rational numbers.  For this scheme, $S$ is asked to be large.

\subsubsection{Phase~1}
\begin{figure}
\[\ba{c}
\includegraphics[width=7.5cm]{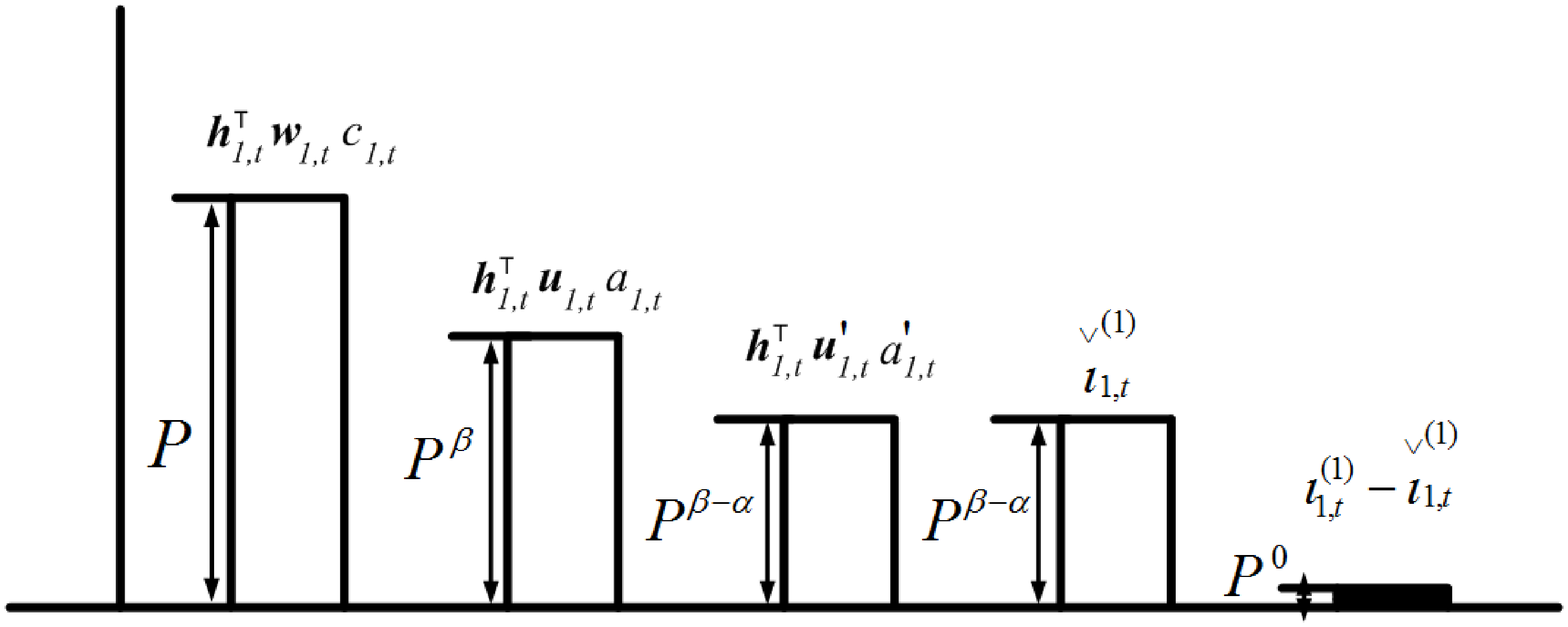}\\
\includegraphics[width=7.5cm]{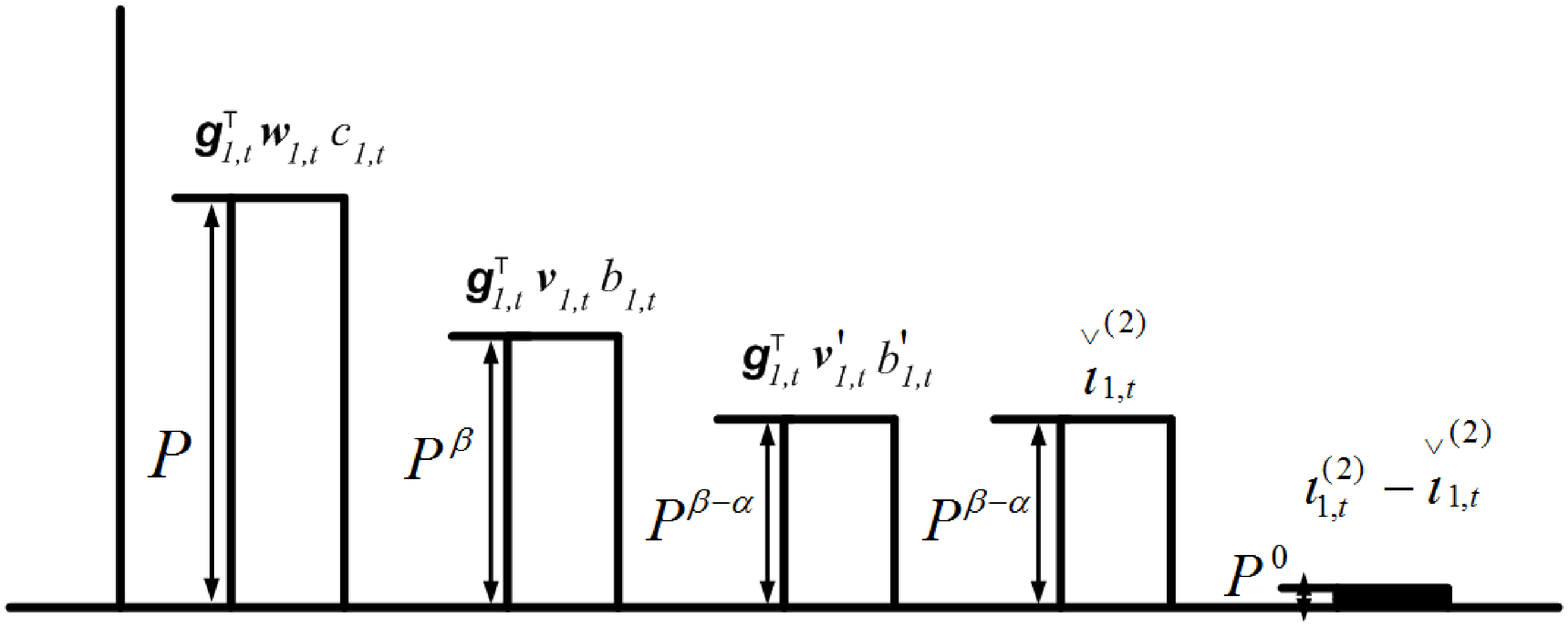}\ea\]
\caption{Received power levels at user~1 (upper) and user~2 (lower): phase~1 of scheme $\Xc_1$. }
\label{fig:MixedX1P1R1}
\end{figure}

During phase~1 ($T_1$ channel uses), the transmitter sends
\begin{align} \label{eq:TxX1Ph1}
\xv_{1,t} =\wv_{1,t} c_{1,t}+\uv_{1,t} a_{1,t} + \uv^{'}_{1,t} a^{'}_{1,t}+\vv_{1,t} b_{1,t}+\vv^{'}_{1,t} b^{'}_{1,t}, \end{align}
with power and rate set as
\begin{equation}\label{eq:RPowerX1Ph1}
\begin{array}{ccc}
P^{(c)}_1 \doteq P, & P^{(a)}_1 \doteq P^{(b)}_1\doteq P^{\beta}, & P^{(a')}_1 \doteq P^{(b')}_1 \doteq P^{\beta-\alpha}\\
r^{(c)}_1  = 1-\beta, &r^{(a)}_1 =r^{(b)}_1 = \beta, & r^{(a')}_1 = r^{(b')}_1 =\beta-\alpha. \end{array}
\end{equation}
The received signals take the form
\begin{align}
  y^{(1)}_{1,t}&= \underbrace{\hv^\T_{1,t} \wv_{1,t} c_{1,t}}_{P}+\underbrace{\hv^\T_{1,t} \uv_{1,t} a_{1,t}}_{P^{\beta}} +\underbrace{\hv^\T_{1,t} \uv^{'}_{1,t} a^{'}_{1,t}}_{P^{\beta-\alpha}} \nonumber\\
  & \!+\!\overbrace{\underbrace{\check{\hv}^\T_{1,t} ( \vv_{1,t} b_{1,t}+ \vv^{'}_{1,t} b^{'}_{1,t})}_{P^{\beta-\alpha}}}^{\check{\iota}^{(1)}_{1,t}} \!+\!  \underbrace{\ddot{\hv}^\T_{1,t} ( \vv_{1,t} b_{1,t}+ \vv^{'}_{1,t} b^{'}_{1,t})}_{P^{0}} \!+\!\underbrace{z^{(1)}_{1,t}}_{P^0}, \label{eq:sch1y1}\\
  y^{(2)}_{1,t}	&= \underbrace{\gv^\T_{1,t} \wv_{1,t} c_{1,t}}_{P}+\underbrace{\gv^\T_{1,t} \vv_{1,t} b_{1,t}}_{P^{\beta}}+\underbrace{\gv^\T_{1,t} \vv^{'}_{1,t} b^{'}_{1,t}}_{P^{\beta-\alpha}}\nonumber\\
 &  \!+\!\overbrace{\underbrace{\check{\gv}^\T_{1,t} ( \uv_{1,t} a_{1,t}\!+\! \uv^{'}_{1,t} a^{'}_{1,t})}_{P^{\beta-\alpha}}}^{\check{\iota}^{(2)}_{1,t}} \!+\! \underbrace{\ddot{\gv}^\T_{1,t} ( \uv_{1,t} a_{1,t}\!+\! \uv^{'}_{1,t} a^{'}_{1,t})}_{P^{0}} \!+\!\underbrace{z^{(2)}_{1,t}}_{P^0}, \label{eq:sch1y2}
\end{align}
where under each term we noted the order of the summand's average power, and where
\begin{align}  \label{eq:barcPower1} \E|\check{\iota}^{(1)}_{1,t}|^2\!&=\! \E|\check{\hv}^\T_{1,t} \vv_{1,t} b_{1,t}|^2+ \E|\check{\hv}^\T_{1,t} \vv^{'}_{1,t} b^{'}_{1,t}|^2\nonumber\\
&= \! \E|(\tilde{\hv}^\T_{1,t}\!-\!\ddot{\hv}^\T_{1,t} )\vv_{1,t} b_{1,t}|^2\!+\! \E|\check{\hv}^\T_{1,t} \vv^{'}_{1,t} b^{'}_{1,t}|^2 \!\doteq\!  P^{\beta-\alpha},\nonumber\\
\E|\check{\iota}^{(2)}_{1,t}|^2\!&= \! \E|(\tilde{\gv}^\T_{1,t}\!-\!\ddot{\gv}^\T_{1,t} )\uv_{1,t} a_{1,t}|^2\!+\! \E|\check{\gv}^\T_{1,t} \uv^{'}_{1,t} a^{'}_{1,t}|^2 \!\doteq\!  P^{\beta-\alpha},
\end{align}
and
\begin{align}\label{eq:barcPower2}
\E|\iota^{(1)}_{1,t}-\check{\iota}^{(1)}_{1,t}|^2\!&= \! \E|\ddot{\hv}^\T_{1,t} ( \vv_{1,t} b_{1,t}+ \vv^{'}_{1,t} b^{'}_{1,t})|^2\!\doteq\!  P^{0}, \nonumber \\
\E|\iota^{(2)}_{1,t}-\check{\iota}^{(2)}_{1,t}|^2\!&= \! \E|\ddot{\gv}^\T_{1,t} ( \uv_{1,t} a_{1,t}+ \uv^{'}_{1,t} a^{'}_{1,t})|^2\!\doteq\!  P^{0}.
\end{align}
Fig.~\ref{fig:MixedX1P1R1} provides a graphical illustration of the received power levels at user~1 and user~2 during phase~1 of scheme $\Xc_1$.

At this point, based on the received signals in~\eqref{eq:sch1y1},\eqref{eq:sch1y2}, each user decodes $c_{1,t}$ by treating the other signals as noise.
The details regarding the achievability of $r^{(c)}_1  = 1-\beta$ can be found in the Appendix. After decoding $c_{1,t}$, user~1 removes $\hv^\T_{1,t} \wv_{1,t} c_{1,t}$ from $y^{(1)}_{1,t}$, while user~2 removes $\gv^\T_{1,t} \wv_{1,t} c_{1,t}$ from $y^{(2)}_{1,t}$.
Then, at the end of the first phase, the transmitter uses its partial knowledge of delayed CSIT to reconstruct $\{\check{\iota}^{(2)}_{1,t}, \check{\iota}^{(1)}_{1,t}\}_{t=1}^{T_1}$ (cf.\eqref{eq:cbarg}), and to quantize each term as
\begin{align} \label{eq:quntisch1}
  \bar{\check{\iota}}^{(2)}_{1,t} = \check{\iota}^{(2)}_{1,t} - \tilde{\iota}^{(2)}_{1,t} , \quad \bar{\check{\iota}}^{(1)}_{1,t} = \check{\iota}^{(1)}_{1,t} -\tilde{\iota}^{(1)}_{1,t} , \quad t=1,2,\cdots, T_1,
\end{align}
where $\bar{\check{\iota}}^{(2)}_{1,t},\bar{\check{\iota}}^{(1)}_{1,t}$ are the quantized delayed estimates of the interference terms, and where $\tilde{\iota}^{(2)}_{1,t},\tilde{\iota}^{(1)}_{1,t}$ are the corresponding quantization errors.
Noting that $\E|\check{\iota}^{(2)}_{1,t}|^2 \doteq P^{\beta-\alpha}, \ \E|\check{\iota}^{(1)}_{1,t}|^2 \doteq P^{\beta-\alpha}$ (cf.~\eqref{eq:barcPower1},\eqref{eq:barcPower2}), we choose a quantization rate that assigns each $\bar{\check{\iota}}^{(2)}_{1,t}$ a total of $(\beta-\alpha)\log P + o(\log P)$ bits, and each $\bar{\check{\iota}}^{(1)}_{1,t}$ a total of $(\beta-\alpha)\log P + o(\log P)$ bits, thus allowing for $\E|\tilde{\iota}^{(2)}_{1,t}|^2 \doteq \E|\tilde{\iota}^{(1)}_{1,t}|^2 \doteq 1$ (see for example~\cite{CT:06}).
At this point, the $2T_1(\beta-\alpha)\log P+o(\log P)$ bits representing $\{\bar{\check{\iota}}^{(2)}_{1,t},\bar{\check{\iota}}^{(1)}_{1,t}\}_{t=1}^{T_1}$, are distributed evenly across the set $\{c_{2,t}\}_{t=1}^{T_2}$ of newly constructed symbols which will be sequentially transmitted during the next (second) phase. This transmission of $\{c_{2,t}\}_{t=1}^{T_2}$ in the next phase, will help each of the users cancel the dominant part of the interference from the other user, and it will also serve as an extra observation (which will in turn enable the creation of a corresponding MIMO channel - see~\eqref{eq:firstMIMO} later on) that allows for decoding of all private information of that same user.

%------------------------
\subsubsection{Phase~$s$, \ $2\leq s\leq S-1$}

Phase~$s$ ($T_s = T_{s-1} \frac{2(\beta-\alpha)}{1-\beta} $ channel uses) is similar to phase~1, with the transmit signal taking the same form as in phase~1 (cf.~\eqref{eq:TxGeneralg},\eqref{eq:TxX1Ph1}), and so do the rates and powers of the symbols (cf.~\eqref{eq:RPowerX1Ph1}), as well as the received signals $y^{(1)}_{s,t},y^{(2)}_{s,t}$ ($ t=1,\cdots,T_s$) (cf.~\eqref{eq:sch1y1},\eqref{eq:sch1y2}).

At the receivers (see~\eqref{eq:sch1y1},\eqref{eq:sch1y2}, corresponding now to phase $s$), each user decodes $c_{s,t}$ by treating the other signals as noise. After decoding $c_{s,t}$, user~1 removes $\hv^\T_{s,t} \wv_{s,t} c_{s,t}$ from $y^{(1)}_{s,t}$, and user~2 removes $\gv^\T_{s,t} \wv_{s,t} c_{s,t}$ from $y^{(2)}_{s,t}$.

At this point, each user goes back one phase and reconstructs, using its knowledge of $\{c_{s,t}\}_{t=1}^{T_s}$, the quantized delayed estimates $\{\bar{\check{\iota}}^{(2)}_{s-1,t},\bar{\check{\iota}}^{(1)}_{s-1,t},\}_{t=1}^{T_{s-1}}$ of all the interference accumulated during the previous phase $s-1$ (cf.\eqref{eq:cbarg},\eqref{eq:quntisch1}).  User~1 then subtracts $\bar{\check{\iota}}^{(1)}_{s-1,t}$ from $y^{(1)}_{s-1,t}$ to remove, up to bounded noise, the interference corresponding to $\check{\iota}^{(1)}_{s-1,t}$.  The same user also employs the estimate $\bar{\check{\iota}}^{(2)}_{s-1,t}$ of $\check{\iota}^{(2)}_{s-1,t}$ as an extra observation which, together with the observation $y^{(1)}_{s-1,t}-\hv^\T_{s-1,t} \wv_{s-1,t} c_{s-1,t} - \bar{\check{\iota}}^{(1)}_{s-1,t}$, allow for decoding of both $a_{s-1,t}$ and $a^{'}_{s-1,t}$.  Specifically user~1, using its knowledge of $\bar{\check{\iota}}^{(2)}_{s-1,t}$, and $y^{(1)}_{s-1,t}-\hv^\T_{s-1,t} \wv_{s-1,t} c_{s-1,t} - \bar{\check{\iota}}^{(1)}_{s-1,t}$, is presented, at this instance, with a $2\times 2$ equivalent MIMO channel of the form
\begin{multline}\label{eq:firstMIMO}
\begin{bmatrix} \!y^{(1)}_{s-1,t}-\hv^\T_{s-1,t}\wv_{s-1,t} c_{s-1,t}\!-\!\bar{\check{\iota}}^{(1)}_{s-1,t}
          \\ \bar{\check{\iota}}^{(2)}_{s-1,t} \!\end{bmatrix}  \\ \!\!=\!\! \begin{bmatrix} \! \hv^\T_{s-1,t} \\ \check{\gv}^\T_{s-1,t} \!\end{bmatrix} \!\!\begin{bmatrix} \! \uv_{s-1,t}  \   \uv^{'}_{s-1,t} \!\end{bmatrix} \!\!\begin{bmatrix} \!a_{s-1,t} \\  a^{'}_{s-1,t} \!\end{bmatrix}
 \!\!+\!\!\! {\begin{bmatrix}  \tilde{z}^{(1)}_{s-1,t}\\
              -\tilde{\iota}^{(2)}_{s-1,t} \!\end{bmatrix}}
\end{multline}
where $\tilde{z}^{(1)}_{s-1,t}$ is the equivalent noise that will be seen to be properly bounded.  As will be argued further in the Appendix, the above MIMO channel allows for decoding of $a_{s-1,t}$ and $a^{'}_{s-1,t}$.

Similar actions are performed by user~2 which uses knowledge of $\bar{\check{\iota}}^{(1)}_{s-1,t}$ and $y^{(2)}_{s-1,t}-\gv^\T_{s,t} \wv_{s,t} c_{s,t} - \bar{\check{\iota}}^{(2)}_{s-1,t}$ to decode both $b_{s-1,t}$ and $b^{'}_{s-1,t}$ (see the Appendix for more details on the achievability of the mentioned rates).

As before, after the end of phase $s$, the transmitter uses its imperfect knowledge of delayed CSIT to reconstruct $\{\check{\iota}^{(2)}_{s,t}, \check{\iota}^{(1)}_{s,t}\}_{t=1}^{T_s}$, and quantize each term to $\bar{\check{\iota}}^{(2)}_{s,t},\bar{\check{\iota}}^{(1)}_{s,t}$ with the same rate as in phase~1 ($(\beta-\alpha)\log P + o(\log P)$ bits for each $\bar{\check{\iota}}^{(2)}_{s,t}$, and $(\beta-\alpha)\log P + o(\log P)$ bits for each $\bar{\check{\iota}}^{(1)}_{s,t}$). Finally the accumulated $2T_{s}(\beta-\alpha)\log P+o(\log P)$ bits representing all the quantized values $\{\bar{\check{\iota}}^{(2)}_{s,t},\bar{\check{\iota}}^{(1)}_{s,t}\}_{t=1}^{T_s}$, are distributed evenly across the set $\{c_{s+1,t}\}_{t=1}^{T_{s+1}}$, the elements of which will be sequentially transmitted in the next phase (phase $s+1$).

%------------------------
\subsubsection{Phase~$S$}
During the last phase ($T_S = T_{S-1}\frac{2(\beta-\alpha)}{1-\alpha} $ channel uses), the transmitter sends
\begin{equation}\label{eq:TxX1PhS}\xv_{S,t} =\wv_{S,t} c_{S,t}+\uv_{S,t} a_{S,t} + \vv_{S,t} b_{S,t}\end{equation}
with power and rates set as
\begin{equation}\label{eq:RPowerX1PhS}
\begin{array}{ll}
P^{(c)}_S \doteq P, & r^{(c)}_S  = 1-\alpha \\
P^{(a)}_S \doteq P^{\alpha} , & r^{(a)}_S  = \alpha\\
P^{(b)}_S \doteq P^{\alpha} , &  r^{(b)}_S =\alpha,
\end{array} \end{equation}
resulting in received signals of the form
\begin{align}
  y^{(1)}_{S,t}\!\!	&=\!\! \underbrace{\hv^\T_{S,t} \wv_{S,t} c_{S,t}}_{P} \!+\!\underbrace{\hv^\T_{S,t} \uv_{S,t} a_{S,t}}_{P^{\alpha}} \! +\!\underbrace{\tilde{\hv}^\T_{S,t} \vv_{S,t} b_{S,t}}_{P^{0}}\!+\!\underbrace{z^{(1)}_{S,t}}_{P^0}, \label{eq:sch1PhSy1} \\
  y^{(2)}_{S,t}\!\!&= \!\!\underbrace{\gv^\T_{S,t} \wv_{S,t} c_{S,t}}_{P} +\underbrace{\tilde{\gv}^\T_{S,t}\uv_{S,t} a_{S,t}}_{P^{0}} \! +\!\underbrace{\gv^\T_{S,t} \vv_{S,t} b_{S,t}}_{P^{\alpha}}+\underbrace{z^{(2)}_{S,t}}_{P^0}, \label{eq:sch1PhSy2}
\end{align}
($ t\!=\!1,\!\cdots\!,T_S$).

As before, both receivers decode $c_{S,t}$ by treating all other signals as noise.  Consequently user~1 removes $\hv^\T_{S,t} \wv_{S,t} c_{S,t}$ from $y^{(1)}_{S,t}$ and decodes $a_{S,t}$, and user~2 removes $\gv^\T_{S,t} \wv_{S,t} c_{S,t}$ from $y^{(2)}_{S,t}$ and decodes $b_{S,t}$. Finally each user goes back one phase and, using knowledge of $\{c_{S,t}\}_{t=1}^{T_{S}}$, reconstructs $\{\bar{\check{\iota}}^{(2)}_{S-1,t},\bar{\check{\iota}}^{(1)}_{S-1,t}\}_{t=1}^{T_{S-1}}$, which in turn allows for decoding of $a_{S-1,t}$ and $a^{'}_{S-1,t}$ at user~1, and of $b_{S-1,t}$ and $b^{'}_{S-1,t}$ at user~2, all as described in the previous phases (see Appendix~\ref{sec:Achievable} for more details).

Table~\ref{tab:x1summary} summarizes the parameters of scheme $\Xc_1$.  In the table, the use of symbol $\bot$ is meant to indicate precoding that is orthogonal to the current channel estimate (else the precoder is generated pseudo-randomly). The last row indicates the prelog factor of the quantization rate.
%---------------------
\begin{table}
\caption{Summary of scheme $\Xc_1$.}
\begin{center}
\begin{tabular}{|c|c|c|c|}
  \hline
                &Phase 1   & Ph.$s$ $(2\!\leq\! s \!\leq\! S\!-\!1)$& Phase $S$\\
   \hline
   Duration     &$T_1$ & $T_1\xi^{s-1}$ & $T_1 \xi^{S-2}\zeta$ \\
   \hline
   $r^{(a)}$    &$\beta$ & $\beta$ & $\alpha$ \\
    \hline
   $r^{(a')}$   &$\beta-\alpha$  & $\beta-\alpha$ & - \\
    \hline
   $r^{(b)}$    &$\beta$  & $\beta$ & $\alpha$ \\
    \hline
   $r^{(b')}$   &$\beta-\alpha$  & $\beta-\alpha$ & - \\
    \hline
	 $r^{(c)}$    &$1-\beta$  & $1-\beta$ &  $1-\alpha$ \\
    \hline
   $P^{(a)}\bot$ &$P^{\beta}$  & $P^{\beta}$ & $P^{\alpha}$ \\
    \hline
   $P^{(a')}$    &$P^{\beta-\alpha}$ & $P^{\beta-\alpha}$ & - \\
    \hline
	 $P^{(b)}\bot$ &$P^{\beta}$ & $P^{\beta}$ & $P^{\alpha}$ \\
    \hline
   $P^{(a')}$    &$P^{\beta-\alpha}$ & $P^{\beta-\alpha}$ & - \\
    \hline
   $P^{(c)}$     &$P$  & $P$ & $P$ \\
    \hline
   Quant.        &$2(\beta-\alpha) $ & $2(\beta-\alpha)$ & $0$ \\
    \hline
\end{tabular}
\end{center}
\label{tab:x1summary}
\end{table}

\paragraph{DoF calculation for scheme $\Xc_1$}
We proceed to add up the total amount of information transmitted during this scheme.

In accordance to the declared pre-log factors $r_s^{(a)},r_s^{(a^{'})},r_s^{(b)},r_s^{(b^{'})},$ given the phase durations (see Table~\ref{tab:x1summary}), and after splitting the common information $\{c_{1,t}\}^{T_1}_{t=1}$ evenly between the two users, we have the two DoF values given by
\begin{align}
d_1&=d_2=\frac{T_1(\frac{1-\beta}{2}+2\beta-\alpha)+\sum^{S-1}_{i=2}T_i(2\beta-\alpha)+T_S\alpha}{\sum^{S}_{i=1}T_i} \nonumber\\
&=2\beta-\alpha+ \frac{ T_1\frac{1-\beta}{2}+2T_S (\alpha-\beta)}{\sum^{S}_{i=1}T_i} \nonumber\\
&=2\beta-\alpha+ \frac{ T_1\frac{1-\beta}{2}+2T_1\xi^{S-2}\zeta (\alpha-\beta)}{T_1(\sum^{S-2}_{i=0}\xi^{i} )+T_1\xi^{S-2}\zeta}.
\end{align}
Considering the case $0<\beta<\frac{1+2\alpha}{3}$ ($0<\xi<1$, see \eqref{eq:X1T}), we see that
\begin{align}
d_1&=d_2=2\beta-\alpha+ \frac{ \frac{1-\beta}{2}+2\xi^{S-2}\zeta (\alpha-\beta)}{\frac{1-\xi^{S-1}}{1-\xi}+\xi^{S-2}\zeta}\nonumber\\
&= 2\beta-\alpha+ \frac{ \frac{1-\beta}{2}+2\xi^{S-2}\zeta (\alpha-\beta)}{\frac{1}{1-\xi}+\xi^{S-2}(\zeta-\frac{\xi}{1-\xi}) },\nonumber
\end{align}
which, for asymptotically high $S$, gives that
\begin{align}
d_1&=d_2= 2\beta-\alpha+  \frac{(1-\beta)(1-\xi)}{2} \nonumber\\
&= 2\beta-\alpha+ \frac{1-3\beta+2\alpha}{2}=\frac{1+\beta}{2}.
\end{align}
Similarly for the case where $\beta=\frac{1+2\alpha}{3}$ ($\xi=1$), we have that
\begin{align}
d_1&=d_2= 2\beta-\alpha+  \frac{ \frac{1-\beta}{2}+2\zeta (\alpha-\beta)}{S-1+\zeta}  \nonumber\end{align}
which, for asymptotically high $S$, gives that
\begin{align}d_1&=d_2= 2\beta-\alpha =\frac{2+\alpha}{3}. \end{align}
Furthermore when $\beta>\frac{1+2\alpha}{3}$ ($\xi>1$), we get that
\begin{align}
d_1 &=d_2= 2\beta-\alpha+\frac{ \frac{1-\beta}{2}+2\xi^{S-2}\zeta (\alpha-\beta)}{\frac{1-\xi^{S-1}}{1-\xi}+\xi^{S-2}\zeta}  \nonumber\end{align}
which, for asymptotically high $S$, gives
\begin{align}d_1&=d_2=   2\beta-\alpha+ \frac{2\zeta (\alpha-\beta)}{\zeta-\frac{\xi}{1-\xi}}  \nonumber\\
&=2\beta-\alpha+ \frac{2(1-3\beta+2\alpha)}{3} =\frac{2+\alpha}{3}.
\end{align}

We can now conclude that scheme $\Xc_1$ achieves the stated DoF pair $C_1=(\frac{1+\beta^{''}}{2},\frac{1+\beta^{''}}{2})$.

\begin{remark}
The observant reader may have noticed that the combination of superposition coding, successive cancelation and power allocation, was calibrated so that, at any fixed receiver, the interfering symbols are received at an equal and bounded power which changes with the quality of current CSIT, and where this interference power is regulated so that, on the one hand, it is sufficiently large to be used as an extra observation by the other user, while on the other hand this interference power remains sufficiently small so that the interference can be reconstructed sufficiently well using bounded quantization rate and imperfect delayed CSIT.  This reconstructed interference is communicated during the next phase, at the expense of having to reduce the amount of new information sent during this next phase.  The relationship, between the amount of interference and new information, is combinatorially optimized by the choice of the phase durations that follow a geometric progression governed by the values of $\alpha$ and $\beta$.
\end{remark}

%==================================
%==================================
%==================================
%==================================
\subsection{Scheme $\Xc_2$ achieving $(\alpha, 1)$ and $(1, \alpha)$: (any $\alpha,\beta$)}
The current scheme applies to the general case of any $\alpha,\beta\in [0,1]$.  This is a simpler scheme and it consists of a single channel use\footnote{We will henceforth maintain the same notation as before, but for simplicity we will remove the phase and time index.} ($S = 1, T_1 = 1$) during which the transmitter sends
\[\xv =\wv c+ \uv a + \vv b,\]
where $\uv$ is orthogonal to the current CSIT estimate $\hat{\gv}$, where $\vv$ is orthogonal to $\hat{\hv}$, and where the power and rate are set as
\begin{equation}\label{eq:ratePowerPhaseSX4}
\begin{array}{ll}
P^{(c)} \doteq P, & r^{(c)}  = 1-\alpha \\
P^{(a)} \doteq P^{\alpha} , & r^{(a)}  = \alpha\\
P^{(b)} \doteq P^{\alpha} , & r^{(b)} =\alpha,
\end{array} \end{equation}
resulting in received signals of the form
\begin{subequations}
\begin{align}
  y^{(1)} &= \hv^\T \xv + z^{(1)}= \underbrace{\hv^\T \wv c}_{P} +\underbrace{\hv^\T \uv a}_{P^{\alpha}} +\underbrace{\tilde{\hv}^\T \vv b}_{P^0}+\underbrace{z^{(1)}}_{P^0},  \nonumber\\ 
  y^{(2)} &= \gv^\T \xv + z^{(2)}= \underbrace{\gv^\T \wv c}_{P} +\underbrace{\tilde{\gv}^\T \uv a}_{P^0} +\underbrace{\gv^\T \vv b}_{P^{\alpha}}+\underbrace{z^{(2)}}_{P^0}. \nonumber
\end{align}
\end{subequations}

After transmission, both receivers decode $c$ by treating the other signals as noise, and then proceed to remove $\hv^\T \wv c$ and $\gv^\T \wv c$, respectively, from their received signals, to get
\beq\label{eq:yMinusInterf1} y^{'(1)} = y^{(1)} - \hv^\T \wv c =  \hv^\T \uv a + \tilde{\hv}^\T \vv b + z^{(1)}=  \hv^\T \uv a + z^{'(1)}\eeq
\[y^{'(2)} = y^{(2)} - \gv^\T \wv c =  \gv^\T \vv b + \tilde{\gv}^\T \uv a + z^{(2)}=  \gv^\T \vv b + z^{'(2)}.\]
The fact that $\E|\tilde{\hv}^\T \vv b |^2 \doteq \E |\tilde{\gv}^\T \uv a |^2 \doteq P^0$, allows for decoding of $a$ and $b$.
Finally, the DoF point $(d_1=\alpha, \quad d_2=1)$ can be achieved by associating $c$ to information intended entirely for the second user, while the DoF point $(d_1=1, \quad d_2=\alpha)$ can be achieved by associating $c$ to information intended entirely for the first user.
The details for the achievability of $r^{(a)},r^{(b)},r^{(c)}$ follow closely the exposition of the details of the previous scheme, as these details are shown in the Appendix.

%==================================================
\section{Conclusions} \label{sec:conclu}

This work provided analysis and novel communication schemes for the setting of the two-user MISO BC with imperfect delayed and imperfect current CSIT.  The results reveal that imperfect delayed CSIT can be as useful as perfect delayed CSIT, as well as provide insight on when it is worth improving CSIT quality.

\section{Appendix - Details of achievability proof} \label{sec:Achievable}

We will here focus on achievability details for scheme $\Xc_1$.  The clarifications of the details carry over easily to the other scheme.

Regarding $r^{(c)}_s$ ($1\leq s \leq S-1$, see~\eqref{eq:RPowerX1Ph1}), we recall that during phase~$s$, both users decode $c_{s,t}$ (from $y^{(1)}_{s,t}, y^{(2)}_{s,t}, t=1,\cdots,T_s$ - see \eqref{eq:sch1y1},\eqref{eq:sch1y2} ) by treating all other signals as noise.
Consequently we note that \begin{align}
I(c_{s,t};y^{(1)}_{s,t},\!\hv_{s,t})\!=\!I(c_{s,t};y^{(2)}_{s,t},\!\gv_{s,t}) \!=\!(1\!-\!\beta)\log\! P\!+\! o(\log\! P),  \nonumber
\end{align}
for large $P$, to get
\begin{align}
r^{(c)}_s \!=\! \frac{1}{\log \! P} \min \{I(c_{s,t};y^{(1)}_{s,t},\hv_{s,t}),I(c_{s,t};y^{(2)}_{s,t},\gv_{s,t})\} \!=\!1\!-\!\beta.\nonumber
\end{align}
Similarly for phase~$S$ (see~\eqref{eq:TxX1PhS}-\eqref{eq:sch1PhSy2}), we note that
\begin{align}
I(c_{S,t};y^{(1)}_{S,t},\!\hv_{S,t})\!=\!\!I(c_{S,t};y^{(2)}_{S,t},\!\gv_{S,t}) \!=\!\!(1\!-\!\alpha)\log\! P\!\!+\! o(\log\! P)  \nonumber
\end{align}
to get
\begin{align}
r^{(c)}_S \!=\! \frac{1}{\log \! P} \min \{I(c_{S,t};y^{(1)}_{S,t},\hv_{S,t}),I(c_{S,t};y^{(2)}_{S,t},\gv_{S,t})\} \!=\! 1\!-\!\alpha.\nonumber
\end{align}

Regarding achievability for $r^{(a)}_s=\beta$, $r^{(a')}_s=\beta-\alpha$, $r^{(b)}_s=\beta$ and $r^{(b')}_s=\beta-\alpha$ ($1\leq s \leq S-1$, see~\eqref{eq:TxX1Ph1},\eqref{eq:RPowerX1Ph1},\eqref{eq:sch1y1},\eqref{eq:sch1y2}), we note that during phase~$s$, both users can decode $c_{s,t}$, and as a result user~1 can remove $\hv^\T_{s,t}\wv_{s,t} c_{s,t}$ from $y^{(1)}_{s,t}$, and user~2 can remove $\gv^\T_{s,t}\wv_{s,t} c_{s,t}$ from $y^{(2)}_{s,t}$ ($t=1,\cdots,T_s$).
Furthermore,  after phase~$s+1$, each user can use its knowledge of $\{c_{s+1,t}\}_{t=1}^{T_{s+1}}$ to reconstruct the quantized delayed estimates $\{\bar{\check{\iota}}^{(2)}_{s,t},\bar{\check{\iota}}^{(1)}_{s,t},\}_{t=1}^{T_s}$ of all the interference accumulated during phase $s$.
As a result, corresponding to phase~$s$, user~1 is presented with $T_s$ linearly independent $2\times 2$ equivalent MIMO channels of the form
\begin{align}
\begin{bmatrix} \!y^{(1)}_{s,t}-\hv^\T_{s,t}\wv_{s,t} c_{s,t}\!-\!\bar{\check{\iota}}^{(1)}_{s,t}
          \\ \bar{\check{\iota}}^{(2)}_{s,t} \!\end{bmatrix}   \!\!=\!\! \begin{bmatrix} \! \hv^\T_{s,t} \\ \check{\gv}^\T_{s,t} \!\end{bmatrix} \!\!\begin{bmatrix} \! \uv_{s,t}  \   \uv^{'}_{s,t} \!\end{bmatrix} \!\!\begin{bmatrix} \!a_{s,t} \\  a^{'}_{s,t} \!\end{bmatrix}
 \!\!+\!\!\! {\begin{bmatrix}  \tilde{z}^{(1)}_{s,t}\\
              -\tilde{\iota}^{(2)}_{s,t} \!\end{bmatrix}} \nonumber
\end{align}
$t=1,\cdots,T_s$, where $\tilde{z}^{(1)}_{s,t}= \ddot{\hv}^\T_{s,t} ( \vv_{s,t} b_{s,t}+ \vv^{'}_{s,t} b^{'}_{s,t})+ z^{(1)}_{s,t} + \tilde{\iota}^{(1)}_{s,t}$.  We note that $\E|\ddot{\hv}^\T_{s,t} ( \vv_{s,t} b_{s,t}+ \vv^{'}_{s,t} b^{'}_{s,t})|^2\doteq P^{0}$ (see \eqref{eq:RPowerX1Ph1},\eqref{eq:sch1y1}). The fact that the rate associated to $\{c_{s+1,t}\}_{t=1}^{T_{s+1}}$, matches the quantization rate for $\{\bar{\check{\iota}}^{(2)}_{s,t}, \bar{\check{\iota}}^{(1)}_{s,t}\}_{t=1}^{T_s}$, allows for a bounded variance of the equivalent noise $\tilde{\iota}^{(2)}_{s,t}$ and $\tilde{\iota}^{(1)}_{s,t}$, and in turn allows for decoding of $\{a_{s,t},a^{'}_{s,t}\}_{t=1}^{T_s}$ at a rate corresponding to $r^{(a)}_s=\beta$ and $r^{(a^{'})}_s=\beta-\alpha$.
Similarly user~2 is presented with $T_s$ linearly independent $2\times 2$ MIMO channels of the form
\begin{align}
	\begin{bmatrix} \! \bar{\check{\iota}}^{(1)}_{s,t}
          \\ y^{(2)}_{s,t}-\gv^\T_{s,t}\wv_{s,t} c_{s,t}\!-\! \bar{\check{\iota}}^{(2)}_{s,t} \!\end{bmatrix} \!\!=\!\! \begin{bmatrix} \! \check{\hv}^\T_{s,t} \\ \gv^\T_{s,t} \!\end{bmatrix} \!\!\begin{bmatrix} \! \vv_{s,t}  \   \vv^{'}_{s,t} \!\end{bmatrix} \!\!\begin{bmatrix} \!b_{s,t} \\  b^{'}_{s,t} \!\end{bmatrix}
 \!+\! {\begin{bmatrix} \! -\tilde{\iota}^{(1)}_{s,t}\\
              \tilde{z}^{(2)}_{s,t} \end{bmatrix}} \nonumber
\end{align}
$t=1,\cdots,T_s$, where $\tilde{z}^{(2)}_{s,t}= \ddot{\gv}^\T_{s,t} ( \uv_{s,t} a_{s,t}+ \uv^{'}_{s,t} a^{'}_{s,t})+ z^{(2)}_{s,t} + \tilde{\iota}^{(2)}_{s,t} $, and where $\E|\ddot{\gv}^\T_{s,t} ( \uv_{s,t} a_{s,t}+ \uv^{'}_{s,t} a^{'}_{s,t})|^2\doteq P^{0}$, $\E|\tilde{z}^{(2)}_{s,t}|^2\doteq P^{0}$, $\E|\tilde{\iota}^{(1)}_{s,t}|^2\doteq P^{0}$, thus allowing for decoding of $\{b_{s,t},b^{'}_{s,t}\}_{t=1}^{T_s}$ at rates corresponding to $r^{(b)}_s=\beta$ and $r^{(b^{'})}_s=\beta-\alpha$.

Regarding achievability for $r^{(a)}_S=\alpha$ and $r^{(b)}_S=\alpha$ (see~\eqref{eq:TxX1PhS},\eqref{eq:RPowerX1PhS},\eqref{eq:sch1PhSy1},\eqref{eq:sch1PhSy2}), we note that, after decoding $c_{S,t}$, user~1 can remove $\hv^\T_{S,t} \wv_{S,t} c_{S,t}$ from $ y^{(1)}_{S,t}$, and user~2 can remove $\gv^\T_{S,t} \wv_{S,t} c_{S,t}$ from $y^{(2)}_{S,t}$, ($t=1,\cdots,T_S$).  Consequently during this phase, user~1 sees $T_S$ linearly independent SISO channels of the form
\begin{align}
\tilde{y}^{(1)}_{S,t}\!\defeq\! y^{(1)}_{S,t}\!-\!\hv^\T_{S,t} \wv_{S,t} c_{S,t}\!=\! \hv^\T_{S,t} \uv_{S,t} a_{S,t} \!+\!\tilde{\hv}^\T_{S,t} \vv_{S,t} b_{S,t}\!+\!z^{(1)}_{S,t} \nonumber
\end{align}
($t=1,\cdots,T_S$) which can be readily shown to support $r^{(a)}_S=\alpha$.  A similar argument gives achievability for $r^{(b)}_S=\alpha$.
\hfill $\Box$

%======================================
%\bibliographystyle{IEEEtran}
%\bibliography{IEEEabrv,mixedcsit_refs}
% Generated by IEEEtran.bst, version: 1.13 (2008/09/30)

\end{document}